\documentclass[journal=aamick,manuscript=article,layout=traditional]{achemso}
\setkeys{acs}{maxauthors=6,etalmode=truncate}

\usepackage{natbib}
\usepackage{graphicx}

\usepackage{color,soul}

\title{Mesoscopic simulations of the \emph{in situ} NMR spectra of porous carbon based supercapacitors: Electronic structure and adsorbent reorganisation effects \normalsize}

\author{Anagha~Sasikumar}
\affiliation[1]{CIRIMAT, Universit\'e de Toulouse, CNRS, B\^at. CIRIMAT, 118, route de Narbonne 31062 Toulouse cedex 9, France}
\alsoaffiliation[2]{R\'eseau sur le Stockage \'Electrochimique de l'\'Energie (RS2E), F\'ed\'eration de Recherche CNRS 3459, HUB de l'\'Energie, Rue Baudelocque,  80039 Amiens, France}
\author{Anouar~Belhboub}
\affiliation[1]{CIRIMAT, Universit\'e de Toulouse, CNRS, B\^at. CIRIMAT, 118, route de Narbonne 31062 Toulouse cedex 9, France}
\alsoaffiliation[2]{R\'eseau sur le Stockage \'Electrochimique de l'\'Energie (RS2E), F\'ed\'eration de Recherche CNRS 3459, HUB de l'\'Energie, Rue Baudelocque,  80039 Amiens, France}
\alsoaffiliation[3]{École Centrale Casablanca, Ville verte, Bouskoura, 27182, Casablanca, Morocco}
\author{Camille~Bacon}
\affiliation[4]{Sorbonne Universit\'e, CNRS, Physico-Chimie des \'Electrolytes et Nanosyst\`emes Interfaciaux, F-75005 Paris, France}
\author{Alexander~C.~Forse}
\affiliation[5]{Department of Chemistry, University of Cambridge, Lensfield Road, Cambridge, CB2 1EW, UK}
\author{John~M.~Griffin}
\affiliation[6]{Department of Chemistry, Lancaster University, Lancaster, LA1 4YB, UK}
\author{Clare~P.~Grey}
\affiliation[5]{Department of Chemistry, University of Cambridge, Lensfield Road, Cambridge, CB2 1EW, UK}
\author{Patrice~Simon}
\affiliation[1]{CIRIMAT, Universit\'e de Toulouse, CNRS, B\^at. CIRIMAT, 118, route de Narbonne 31062 Toulouse cedex 9, France}
\alsoaffiliation[2]{R\'eseau sur le Stockage \'Electrochimique de l'\'Energie (RS2E), F\'ed\'eration de Recherche CNRS 3459, HUB de l'\'Energie, Rue Baudelocque,  80039 Amiens, France}
\author{C\'eline~Merlet}
\affiliation[1]{CIRIMAT, Universit\'e de Toulouse, CNRS, B\^at. CIRIMAT, 118, route de Narbonne 31062 Toulouse cedex 9, France}
\alsoaffiliation[2]{R\'eseau sur le Stockage \'Electrochimique de l'\'Energie (RS2E), F\'ed\'eration de Recherche CNRS 3459, HUB de l'\'Energie, Rue Baudelocque,  80039 Amiens, France}
\email{merlet@chimie.ups-tlse.fr}

\date{}

\begin{document}

\newpage

\begin{abstract}
\emph{In situ} NMR spectroscopy is a powerful technique to investigate charge storage mechanisms in carbon-based supercapacitors thanks to its ability to distinguish ionic and molecular species adsorbed in the porous electrodes from those in the bulk electrolyte. The NMR peak corresponding to the adsorbed species shows a clear change of chemical shift as the applied potential difference is varied. This variation in chemical shift is thought to originate from a combination of ion reorganisation in the pores and changes in ring current shifts due to the changes of electronic density in the carbon. While previous Density Functional Theory calculations suggested that the electronic density has a large effect, the relative contributions of these two effects is challenging to untangle. Here, we use mesoscopic simulations to simulate NMR spectra and investigate the relative importance of ion reorganisation and ring currents on the resulting chemical shift. The model is able to predict chemical shifts in good agreement with NMR experiments and indicates that the ring currents are the dominant contribution. A thorough analysis of a specific electrode/electrolyte combination for which detailed NMR experiments have been reported allows us to confirm that local ion reorganisation has a very limited effect but the relative quantities of ions in pores of different sizes, which can change upon charging/discharging, can lead to a significant effect. Our findings suggest that \emph{in situ} NMR spectra of supercapacitors may provide insights into the electronic structure of carbon materials in the future.
\end{abstract}


\section{Introduction}

Electrochemical double-layer capacitors (EDLCs) or supercapacitors have gained importance in recent times as the demand for reliable clean energy sources has increased. They store energy through the formation of an electric double layer at the electrode/electrolyte interface via ion adsorption, resulting in desirable properties such as high charge/discharge rate and long cycle life. Porous carbons are widely used as electrode materials in supercapacitors due to their high surface area for adsorption, high electrical conductivities, low-cost and inert nature~\cite{Simon13,Salanne16,Miao20}. The electrochemical properties of porous carbons are highly dependent on both their local and longer range structure. Despite this, a detailed understanding of the underlying molecular mechanisms and structure-properties relationships that are crucial for improving device performance is lacking. Several experimental and theoretical efforts have been put forward in this aspect although the structural complexity of the porous material raises multiple challenges. 

The development of \emph{in situ} characterization techniques has facilitated the study of supercapacitors under working conditions and has provided valuable insights about charging mechanisms and the structure of the electrode-electrolyte interface. \emph{In situ} electrochemical quartz crystal microbalance (EQCM) has been used to monitor ion fluxes under an applied potential and identify factors influencing it such as the average pore size or the electrolyte nature.~\cite{Levi09,Tsai14,Griffin15} Scattering techniques such as small-angle X-ray scattering (SAXS) or small-angle neutron scattering (SANS) have helped in probing the structure of the carbon electrode as well as ion adsorption and desolvation during the course of charging/discharging.~\cite{Prehal15,Prehal17,Boukhalfa14}

Nuclear Magnetic Resonance (NMR) spectroscopy stands out among these experimental techniques as, owing to its nucleus specificity and sensitivity to local chemical environments, it provides detailed information about the behaviour of electrolyte ions in porous carbon electrodes. The ability to distinguish species that are adsorbed to the electrode surface (in-pore) from those in the bulk and to quantify them has assisted in comprehending the charging mechanisms of supercapacitors.~\cite{Griffin14,Griffin16,Forse18c} Up to now, in-pore electrolyte species are always observed to be shifted to lower frequencies relative to bulk species on account of the presence of a local magnetic field originating from the circulating electrons in the carbon electrode (ring currents). This ring current shift can be approximated by calculating Nucleus Independent Chemical Shifts (NICS). 

Several studies have suggested that the difference in chemical shift between in-pore and bulk species, noted $\Delta \delta$, can be used to determine structural characteristics of the porous carbon.~\cite{Forse14,Forse15b,Xing14,Anderson10} Theoretical studies by Forse \emph{et al.}~\cite{Forse14}, using  Density Functional Theory (DFT) on model aromatic molecules, have explored the effect of structural factors such as carbon pore size and curvature in the carbon structure on the resulting NICS. In general, the closer the ions are to the surface, the more negative the NICS is. Xing \emph{et al.}~\cite{Xing14} have used similar calculations to estimate the average pore size and pore size distribution of activated carbons. Forse~\emph{et al.}~\cite{Forse15b} have shown that the shift between the bulk and in-pore species also brings information regarding the local ordering of the carbon atoms. It is worth noting though that ion and solvent dynamics can also affect the in-pore chemical shift~\cite{Cervini19} making the experimental results more challenging to interpret in depth. 

\emph{In situ}~\cite{Wang13,Griffin14,Griffin16,Forse18c} and \emph{ex situ}~\cite{Deschamps13,Forse15} NMR experiments of organic electrolytes have revealed changes in intensity and chemical shift of the in-pore resonances with charging. On the application of a potential, a number of studies showed that in-pore resonances moved to higher chemical shifts irrespective of the polarity of the electrode. These variations are believed to arise from the reorganisation of the in-pore ions, the alteration of the ring currents in the carbon electrode during charging, or a combination of both. An \emph{in situ} NMR study of NaF aqueous electrolyte in nanoporous carbons has shown a reverse trend, with in-pore resonances moving to lower chemical shifts as the potential increases, for some of the nuclei studied and potentials applied.~\cite{Luo15b} In this case, the influence of the speciation (and in particular hydration number) makes the microscopic picture more complex and this system is not studied here as the focus is on organic electrolytes. DFT calculations on charged aromatic molecules have shown that the NICS varies from negative values close to the surface for neutral molecules to positive values for charged molecules, independently of the positive or negative charge imposed.~\cite{Wang13} This is consistent with the experimental observation in terms of the qualitative variation of the shift but the magnitude of the effect is not understood to date. DFT calculations, while being powerful tools to estimate the NICS, only consider simplistic carbon structures and surface charges and do not allow for ion dynamics in complex mesostructures to be readily taken into account.  

In this study, we simulate \emph{in situ} NMR spectra by using a lattice simulation model~\cite{Merlet15,Anouar19} in order to quantify the chemical shifts and explain the previously observed trends in detail. The mesoscopic model employed here was previously used to model ion diffusion, NMR spectra for neutral carbons and capacitive properties of supercapacitors. It combines inputs from experiments, molecular dynamics simulations (MD) and DFT calculations to predict an NMR spectrum in a very computationally efficient manner. In particular, data from gas adsorption experiments can be used to model carbon particles with the appropriate pore size distribution and data from simulations allow for the assessment of the relative importance of ion reorganisation, included through free energy profiles extracted from molecular simulations, and NICS modification, following DFT calculations, at different potentials. We investigate the NMR spectra obtained experimentally for a range of electrolyte species in a  YP-50F activated carbon and examine the relative importance of free energies and NICS on the resulting chemical shift of in-pore species. The model shows good agreement with experimental results and suggests that the shifts caused by ring currents dominate those that result from ion reorganisations. 

\section{Methods}

\subsection{Overview of the lattice model}

\begin{figure*}[ht!]
\centering
\includegraphics[scale=0.25]{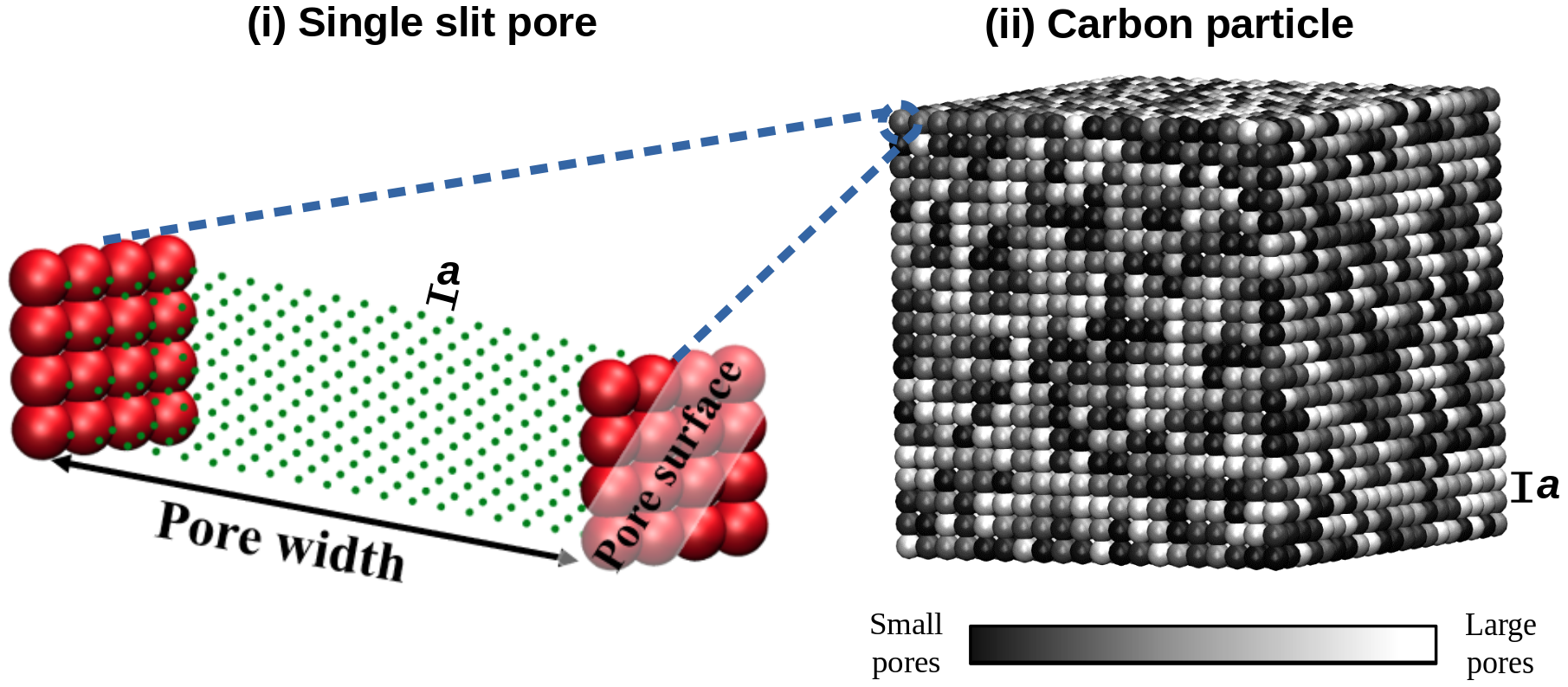}
\caption{Schematic representation of (i) a single slit pore and (ii) a particle in the lattice model. The discrete sites in the lattice are separated by a distance $a$.}
\label{scheme-LM}
\end{figure*}

The lattice simulation method used in this work allows ions diffusing inside porous electrodes as well as their NMR spectra to be modeled.~\cite{Merlet15,Anouar19} Using this method it is possible to simulate the chemical shift difference between the bulk electrolyte species and the in-pore electrolyte species ($\Delta \delta = \delta_{in-pore}-\delta_{bulk}$) at two different scales: (i) a slit pore having a well defined pore size and (ii) a carbon particle with a realistic pore size distribution. A schematic representation of these two models is given in Figure~\ref{scheme-LM}.

The system under consideration is mapped onto a three-dimensional lattice consisting of discrete sites separated by a distance \emph{a}. Each site is assigned an ``accessible" or ``non accessible" character depending on whether it corresponds to a position in the fluid or in the carbon. \emph{$\Delta$t} is the time a particle, here an ion or solvent molecule, takes to diffuse over a distance \emph{a}. The ion diffusion is described by executing kinetic Monte Carlo moves on the lattice. To conduct a simulation, one has to assign:\\
(i) the positions of the inaccessible sites if any;\\
(ii) the free energies for ions (or solvent molecules) on each lattice site,~$E_i$;\\
(iii) the resonant frequencies corresponding to ions (or solvent molecules) on each lattice site,~$w_i$.\\ 
The free energy, $E_i$, accounts for the quantity of adsorbed species at a given site $i$ and is defined using ionic densities from molecular dynamics simulations. The diffusion of ions between lattice sites is then associated with the free energy difference between these sites. The transition probability from a site $i$ to $j$ is given by:
\begin{equation}\label{probability}
P(i,j)= \left\{
\begin{array}{cc}
exp(\frac{-(E_j-E_i)}{k_BT}) & \mathrm{if\ } E_j>E_i \\
1 & \mathrm{if\ } E_j\leq {E_i}
\end{array}
\right.
\end{equation}
Resonant frequencies account for the shielding environment of a site and are defined using DFT calculations directly (for the slit pore model) or through a combination of molecular simulations and DFT results (for the particle model). For an ion at a particular site, the resonance frequency is expressed as the difference between the Larmor frequency of the ion at that site and in the bulk liquid. During the course of its motion, the ion explores sites of different frequencies. For all the nuclei under consideration, the NMR signal is calculated as follows:
\begin{equation}\label{NMR_expression}
     G(t) = \langle e^{i \sum_{n'=1}^{n} \omega_i(n'.\Delta t)\Delta t}  \rangle
\end{equation}
where $<...>$ denotes an average over all ions considered, $\omega_i$ is the Larmor frequency at site $i$, and $n$ designates a given time step.

\subsection{Parametrisation of the model for simulating NMR spectra of ions in a single slit pore}

A slit pore can be described by a lattice with two parallel planes of obstacles representing the pore walls surrounding an accessible volume. In this work, the pore is aligned perpendicular to the $z$-axis. The free energies and chemical shifts are assumed to be invariant in $x$ and $y$ directions, and periodic boundary conditions are applied, thus requiring only a single site in these directions. The considered slit pore has a pore width of 20~\r{A} and was mapped onto a lattice of size $1\times1\times41$ with a lattice parameter $a$ set to 0.5~\r{A}. This pore size was chosen as i) it is representative of a range of activated carbons~\cite{Cervini19,Borchardt17} and ii) it allows for the study of ion reorganisation at the local scale, inside a pore, thanks to the existence of several layers of ions at different distances from the carbon surface. 

\begin{figure}[ht!]
\centering
\includegraphics[scale=0.29]{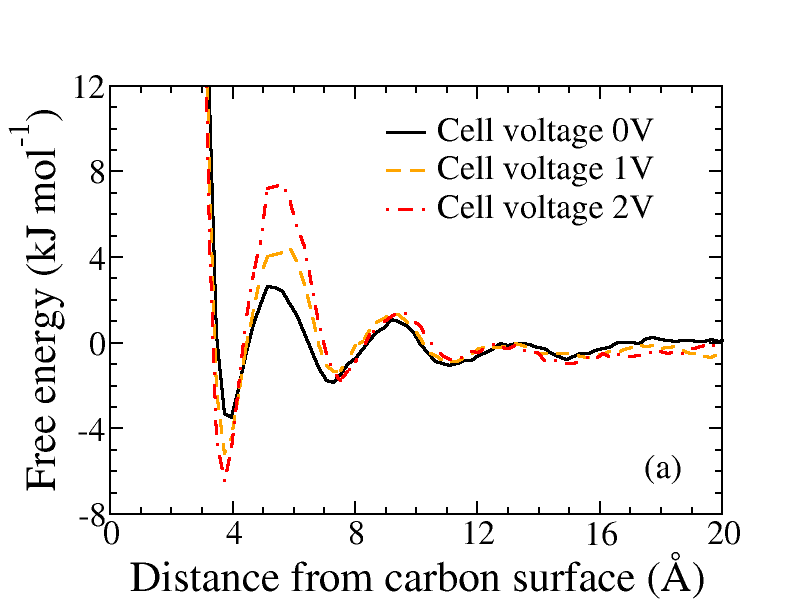}
\includegraphics[scale=0.29]{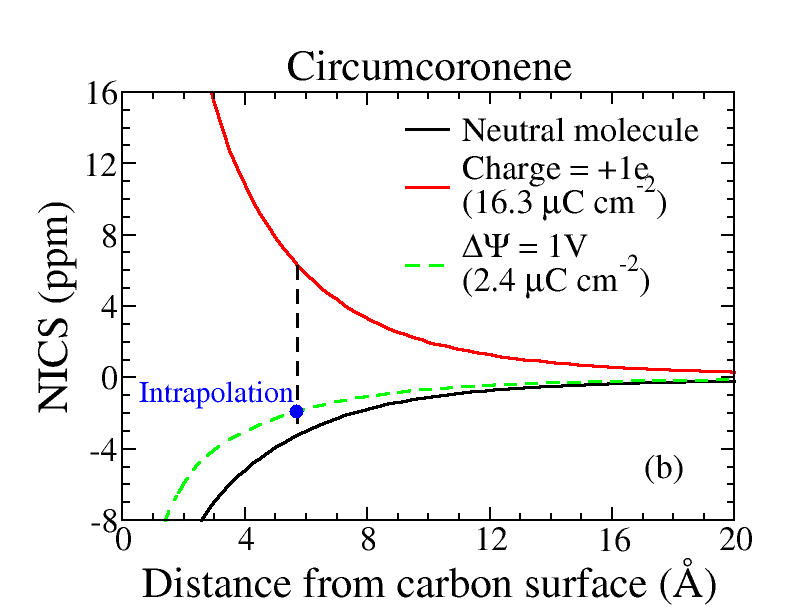}
\caption{(a) Free energy profiles for BF$_4^-$ anions in [BMI][BF$_4$]-ACN (1.5M) at the positive electrode for various applied potential differences. (b) NICS values at various distances obtained from DFT calculations using circumcoronene as model molecule. NICS values for intermediate surface charges are intrapolated from the results of the DFT calculations (see Supplementary Information for details).}
\label{input-slit}
\end{figure}

To assign free energy values to the lattice sites, free energy profiles for anions obtained from MD simulations performed at fixed potentials ranging from -2~V to 2~V in 1-butyl-3-methylimidazolium tetrafluoroborate dissolved in acetonitrile ([BMI][BF\textsubscript{4}] in ACN, 1.5~M)~\cite{Merlet13,Siepmann95,Pounds09,Wang14c} are used. A single electrolyte is studied for this slit pore representation as the chemical shift changes on this local scale are not directly comparable with experimental values. The aim here is to illustrate the qualitative trend on an example as opposed to providing a comparison with measured values. Note that a coarse-grained approach~\cite{Roy10b,Merlet13,Merlet12b} was used to model the molecular species, in particular a single sphere represents the BF$_4^-$ anions (see Supplementary Information). The free energy profiles reflect the adsorption behaviour of ions at various potentials. As can be seen in Figure~\ref{input-slit}a, a decrease in free energy at the minimum around 4~\r{A} indicates a stronger adsorption of anions at the positively-polarised pore surface. On the contrary, the application of a negative potential results in the desorption of anions from the surface (see Supplementary Information).  Each site is then assigned an energy value appropriate to its distance from the site to the pore wall.  

As done in a number of previous works, NICS profiles are determined by performing DFT calculations of NICS in the proximity of aromatic molecules using the Gaussian 09 software.~\cite{Xing14,Forse14,Kilymis20,Moran03,frisch2013,becke1993} Circumcoronene is used as the main model molecule to calculate the chemical shifts as it provides good agreement with the NICS evaluated experimentally for a mesoporous carbon material (CMK-3) and an organic electrolyte.~\cite{Borchardt13,Merlet15} Interestingly, circumcoronene is also the molecule used in the model of Xing~\emph{et~al.}~\cite{Xing14} to determine the average pore size and pore size distribution of porous carbons from NMR experiments. This model provided average pore sizes in good agreement with those obtained from gas adsorption isotherms, when applied to NMR data for aqueous solutions in contact with activated carbons~\cite{Cervini19}. The NICS profiles were estimated for a neutral molecule or ions with a $\pm$1e charge which corresponds to a surface charge of $\pm$16.3~$\mu$C~cm\textsuperscript{-2}. The actual surface charge at the potentials under consideration in real supercapacitors is far less than this.~\cite{Merlet12b} Hence we intrapolate the NICS profiles to obtain chemical shifts for a particular surface charge corresponding to the potential under consideration (Figure~\ref{input-slit}b). Details of the DFT calculations and charges considered for all potentials, extracted from molecular simulations, are given in Supplementary Information.

\subsection{Parametrisation of the model for simulating NMR spectra of ions in a carbon particle}

To represent a realistic carbon particle, a lattice of size $20\times20\times20$ is used, where each lattice site is a slit pore with well defined pore size. This lattice size was previously shown to provide reliable results for NMR spectra calculations.~\cite{Merlet15} The model allows the pore characteristics (pore width and pore surface) as well as the quantity of adsorbed ions (through free energies) and the resonant frequency to be specified. The pore width is defined following the pore size distribution obtained experimentally through gas adsorption studies on the YP-50F activated carbon material.~\cite{Forse17} The pore width is defined as the distance between the centers of the carbon atoms in both the present model and in the \mbox{QSDFT} model used to analyse the gas adsorption experiments. For the pore surface, which corresponds to the area of the flat carbon surfaces surrounding the liquid in the model (see Figure~\ref{scheme-LM}), a log-normal distribution centered around the area of circumcoronene, with a mean value of -0.1 and a standard deviation of 0.25, is used. Different pore surfaces are a way to represent different local graphitization degrees in real porous carbons. The choice of the distribution was again based on the good agreement it provides with the experimental results obtained by Borchardt~\emph{et~al.}~\cite{Borchardt13} and Cervini~\emph{et~al.}~\cite{Cervini19}. Both pore size and pore surface distributions are shown in Supplementary Information, along with results on the use of different pore surfaces. 

The quantity of adsorbed ions in a given slit-pore is obtained from ionic density profiles from MD simulations at various potentials.~\cite{Siepmann95,Pounds09,Wang14c,LAMMPS,MetalWalls,fftool} Several electrolytes are considered for this more realistic model of the carbon particles to allow for a direct comparison with experiments: 1-butyl-3-methylimidazolium tetrafluoroborate in acetonitrile ([BMI][BF$_4$]-ACN, 1.5M), tetraethylphosphonium tetrafluoroborate in acetonitrile ([PEt$_4$][BF$_4$]-ACN, 1.5M) and the neat ionic liquid 1-ethyl-3-methylimidazolium bis(trifluoromethanesulfonyl)imide ([EMI][TFSI]). In MD, the [BMI][BF$_4$]-ACN electrolyte is again represented by a coarse-grained model~\cite{Roy10b,Merlet13,Merlet12b} while the other electrolytes are simulated using all-atom models.~\cite{Canongia-Lopes04,Canongia-Lopes04b,Canongia-Lopes06b,Price01} Molecular simulations details are given in Supplementary Information. Figure~\ref{input-coarse}a shows the integrated densities of BF$_4^-$ in [BMI][BF$_4$] for various pore sizes. The adsorption/desorption of anions at different applied potentials is clearly visible on this plot.
\begin{figure}[ht!]
\centering
\includegraphics[scale=0.29]{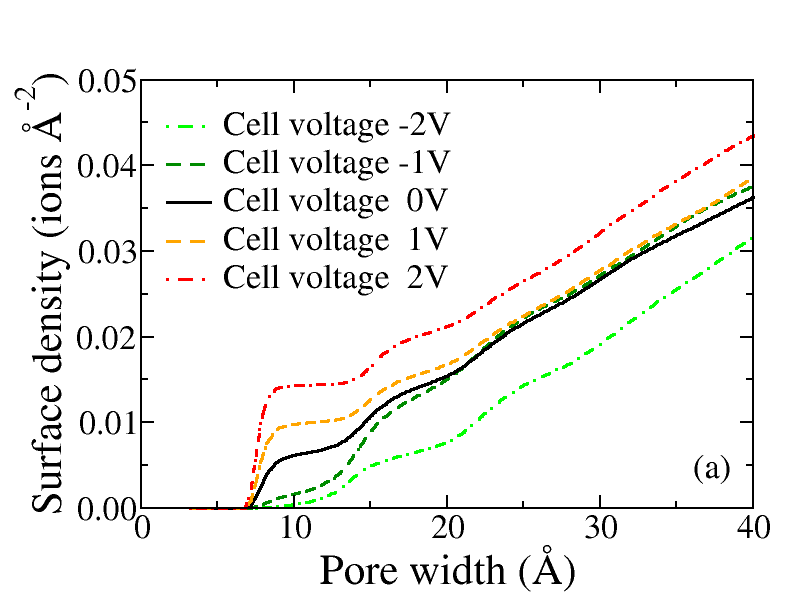}
\includegraphics[scale=0.29]{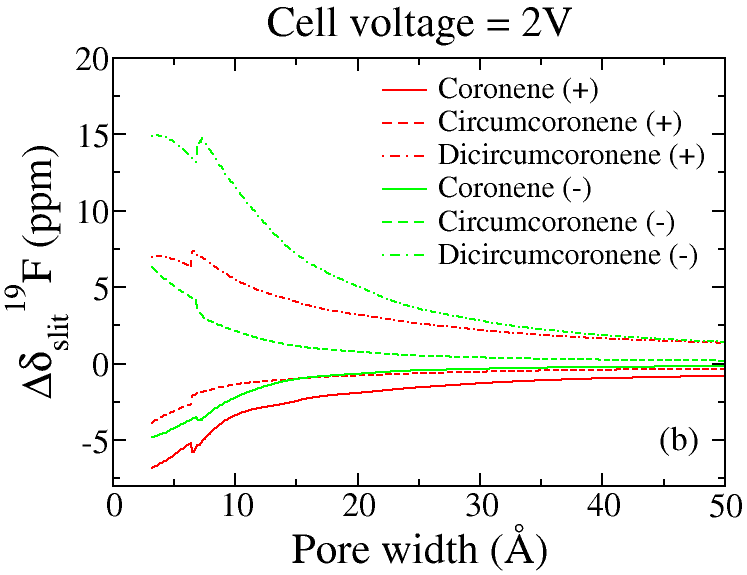}
\caption{Integrated density profiles (a) and average chemical shifts of BF$_4^-$ with respect to the bulk electrolyte ([BMI][BF$_4$] in ACN, 1.5M) (b) as a function of pore width calculated for various potentials and model molecules (coronene, circumcoronene, and dicircumcoronene corresponding to different pore surfaces).}
\label{input-coarse}
\end{figure}

Chemical shifts of an ion inside a pore are calculated using both NICS and free energy profiles. Specifically, the average chemical shift is a weighted average of the NICS profile across the pore (see previous work for details~\cite{Merlet15}). These shifts are largely dependent on the model molecules chosen in the DFT calculations (i.e. coronene, circumcoronene, and dicircumcoronene, corresponding to different pore surfaces) and on the surface charge (Figure~\ref{input-coarse}b). Since a distribution of pore surfaces is considered, the average chemical shifts corresponding to a particular pore surface at an applied potential were determined by intrapolating the NICS profiles calculated using model molecules coronene (36.2~\r{A}$^2$), circumcoronene (98.1~\r{A}$^2$) and dicircumcoronene (191.1~\r{A}$^2$), with the surface charge corresponding to the applied cell voltage.

\subsection{Testing the relative importance of ion reorganisation and ring currents at various potentials}
 
In this study, the aim is to understand the underlying principles behind the voltage-de\-pen\-den\-ce of the chemical shift of in-pore resonances by assessing the influence of ion reorganisation, through free energies, and ring currents, through NICS. The contribution to the global shift of the local ion reorganisation (or the variation in quantities of adsorbed ions in different pores for the particle model) between 0~V and a given applied cell voltage was evaluated by considering i) NICS profiles (or average chemical shifts for the particle model) calculated for a neutral surface and ii) free energies (or integrated ion densities for the particle model) for the respective potentials.  By contrast, the contribution of altered ring currents was assessed by i) considering NICS profiles (or average chemical shifts) calculated for a charged surface and ii) disregarding the change in carbon-ion distance (or quantity of adsorbed ions inside the pore) while charging by considering free energies (or integrated ion densities) at 0~V.

\section{Results and discussion}

\subsection{\emph{In situ} NMR spectra of ions adsorbed in a single slit pore}

\begin{figure}[ht!]
\centering
\includegraphics[scale=0.3]{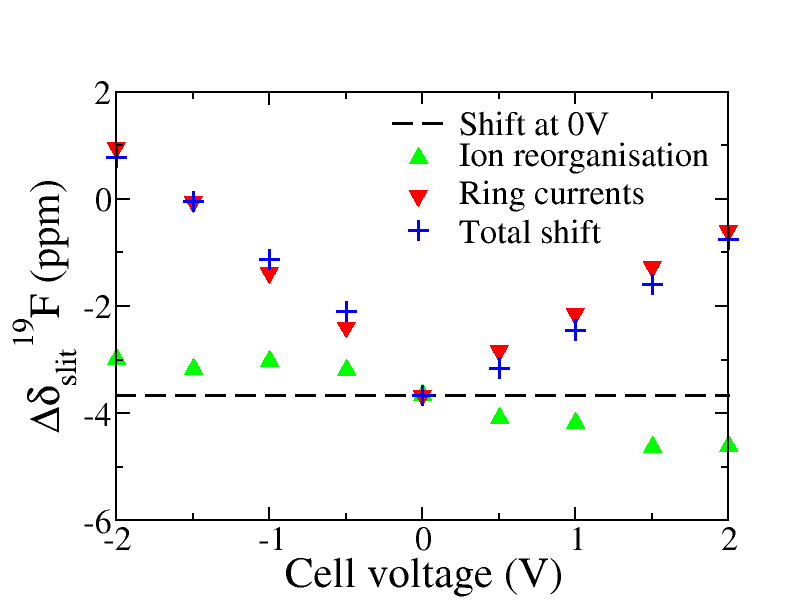}
\caption{Average chemical shift for BF$_4^-$ anions adsorbed inside a 2~nm carbon slit pore relative to the bulk electrolyte and contributions of ion reorganisation and ring currents, obtained through the lattice model for [BMI][BF$_4$] (1.5~M in ACN). The ion reorganisation and ring currents contributions are assessed by fixing the free energies and NICS to their values at 0V in specific calculations.}
\label{results-slit}
\end{figure}

In order to assess the relative importance of (i) ion reorganisation at the electrolyte/electrode interface and (ii) charge dependent ring currents, on the chemical shift variation observed in \emph{in situ} NMR experiments, $^{19}$F NMR spectra were simulated and average chemical shifts for BF$_4^-$ anions confined inside a slit pore with respect to those in the free electrolyte ([BMI][BF$_4$] in ACN, 1.5M), $\Delta\delta_{\rm slit}$ $^{19}$F, were calculated. Average chemical shifts obtained for fixed potential differences between -2~V and 2~V are shown in Figure~\ref{results-slit}. For the total chemical shift, an approximately linear variation to higher frequencies is observed on the application of a potential, independently of the polarity of the pore. For a voltage of 1.5~V across the electrochemical cell, a difference of +2.1~ppm and +3.6~ppm was observed for positive and negative pores respectively compared to 0~V. This is in the same range as that observed experimentally for a similar electrolyte, ([NEt$_4$][BF$_4$] in ACN, 1.5~M), in contact with YP-50F activated carbon, where $^{19}$F chemical shifts changes of +3.7~ppm and +4.4~ppm were observed for positive and negative electrodes respectively~\cite{Griffin14}. While the actual values are different, which is unsurprising due to the crudeness of the slit pore pore model, the larger chemical shift change for negative potentials observed experimentally is reproduced by the lattice model.

When looking at the relative contributions of ring currents and ion reorganisation in the pores, it is very clear that most of the total shift is due to variations in the ring currents. With an applied potential, the ions experience different positions in the pore leading to different ring currents as the NICS depends on the ion-carbon distance.  It is important to note that the ring currents and ion reorganisation contributions are not additive because the average chemical shift results from a convolution of the NICS and free energy profiles which vary differently as a function of the distance to the carbon surface. The chemical shift change due to the NICS is always to more positive chemical shifts irrespective of the electrode polarity as suggested by DFT calculations, owing to the accumulation of electrons/holes on the pore surface rendering paratropic ring current effects. It is worth noting that the absolute changes in NICS are smaller on the positive side than on the negative side.  While the contribution due to the NICS varies proportionally to the applied cell voltage, the ion reorganisation contribution is more constant. The reorganisation leads to a negative chemical shift change for a positively charged pore and to a positive chemical shift change for a negatively charged pore. This is due to a stronger adsorption of BF$_4^-$ anions on a positively charged surface, leading to a shorter carbon-anion distance on average (and a larger ring current effect), and the reverse phenomenon on a negatively charged surface. 

\subsection{\emph{In situ} NMR spectra of ions adsorbed in a carbon particle}

\begin{figure}[ht!]
\centering
\includegraphics[scale=0.3]{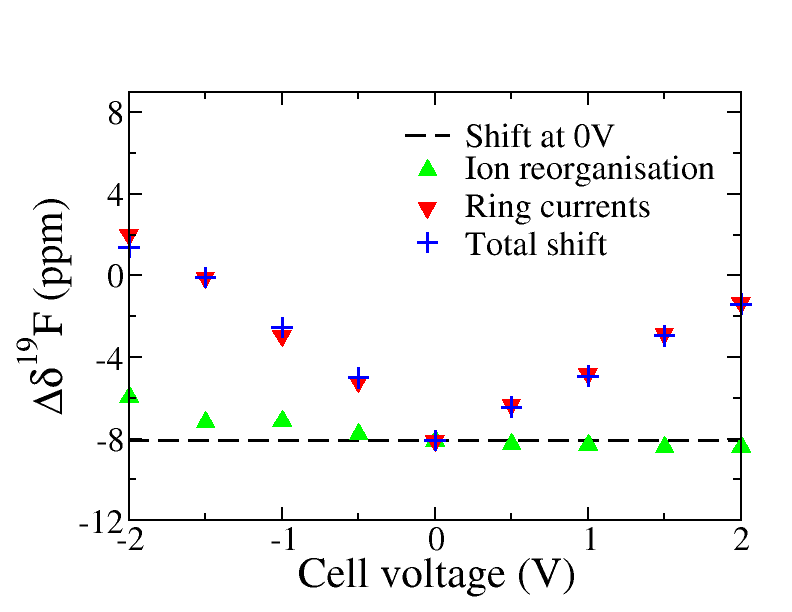}
\caption{Average chemical shift for BF$_4^-$ anions adsorbed inside carbon particle relative to the bulk electrolyte and contributions of ion reorganisation and ring currents, obtained through the lattice model for [BMI][BF$_4$] (1.5M in ACN). The ion reorganisation and ring currents contributions are assessed by fixing the surface densities and average chemical shifts to their values at 0V in specific calculations.}
\label{results-carbon}
\end{figure}

To get closer to experimental conditions, the lattice model was then used with a representation of a carbon particle, which includes its pore size distribution taken as the one of YP-50F, and a pore surface area distribution, which follows a log-normal distribution centered around the area of circumcoronene. The chemical shifts of in-pore BF$_4^-$ anions relative to their bulk counterparts, $\Delta\delta~^{19}$F, obtained in this case ([BMI][BF$_4$] in ACN, 1.5~M) are shown in Figure~\ref{results-carbon}. 
The general trends for this carbon particle model are very similar to the ones observed for the slit pore. In particular, the chemical shift change observed with an applied potential is mainly due to variations in the ring currents, while the effect of ion reorganisation is limited. It is worth noting that the chemical shift changes observed between 0~V and 1.5~V are much larger than for the slit pore, more precisely +5.1~ppm for the positively and +8.0~ppm for the negatively charged particles compared to +2.1~ppm and +3.6~ppm for the slit pore model. The larger $\Delta\delta$ values compared to the slit pore are expected as the pore size distribution of YP-50F (see Supplementary Information) shows mostly pore sizes below 2~nm (the pore size used for the slit pore model) leading to larger chemical shifts for adsorbed ions relative to the bulk. The experimental values for [NEt$_4$][BF$_4$] in ACN, +3.7~ppm and +4.4~ppm, are between the values obtained with the slit-pore and the particle models~\cite{Griffin14}. Nevertheless, as the electrolytes are not exactly the same, it is hard to comment on these differences. As for the slit pore model, the absolute changes in shifts on the positive side are smaller than on the negative side, which is believed to be mainly due to the charge dependence of the NICS, even though for this system both NICS and ion reorganisation are cooperative. 

To investigate the predictive power of the model in more depth, additional simulations were conducted with another organic electrolyte for which experimental results are published, namely [PEt$_4$][BF$_4$] in ACN, and with a neat ionic liquid. While the neat ionic liquid studied in this work, [EMI][TFSI], is not the same as the previously reported on experimental one, [Pyr$_{13}$][TFSI] (1-methyl-1-propylpyrrolidinium
bis(trifluoromethanesulfonyl)imide), they share a common anion and provide lattice simulations results for a system where ionic correlations are not screened by a solvent and the approach, adopted here, of extracting free energy profiles from molecular dynamics simulations could potentially lead to dramatically wrong results. 
\begin{figure}[ht!]
\centering
\includegraphics[scale=0.3]{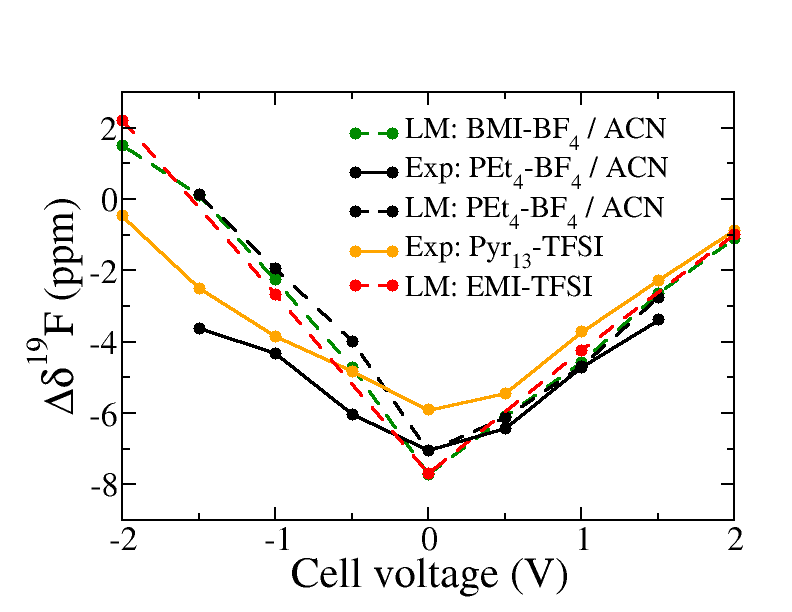}
\caption{Comparison of average chemical shifts for anions adsorbed inside YP-50F carbon particles relative to the bulk electrolyte obtained through experiments~\cite{Griffin15,Forse15} and lattice simulations.}
\label{comp-sim-expt}
\end{figure}
In particular, the quantity of ions adsorbed in small pores could be largely underestimated leading to very small $\Delta\delta$ values compared to experiments. Indeed, if ions are only present in relatively large pores, the average ion-carbon distance is overestimated and the simulated $\Delta\delta$ is underestimated. The comparison between lattice simulations and experimental results for anions is shown in Figure~\ref{comp-sim-expt}. The agreement between lattice simulations and experiments is good, especially for positive potentials. For negative potentials, the lattice simulations seem to overestimate the chemical shift change with potential. It is remarkable that for experiments or simulations, the chemical shift variation with potential, i.e. the slope of the profiles, is very similar for all electrolytes, consistent with a variation mainly due to ring currents. This effect has actually been seen experimentally on a larger range of electrolytes than the one explored here.~\cite{Forse21} According to the results obtained here, the coarse-grained or all-atom representation of the ions and molecules in the molecular simulations do not lead to significant changes. The ion adsorption can still have an effect on the actual chemical shift as can be seen in particular at 0~V. 

\begin{figure}[ht!]
\centering
\includegraphics[scale=0.29]{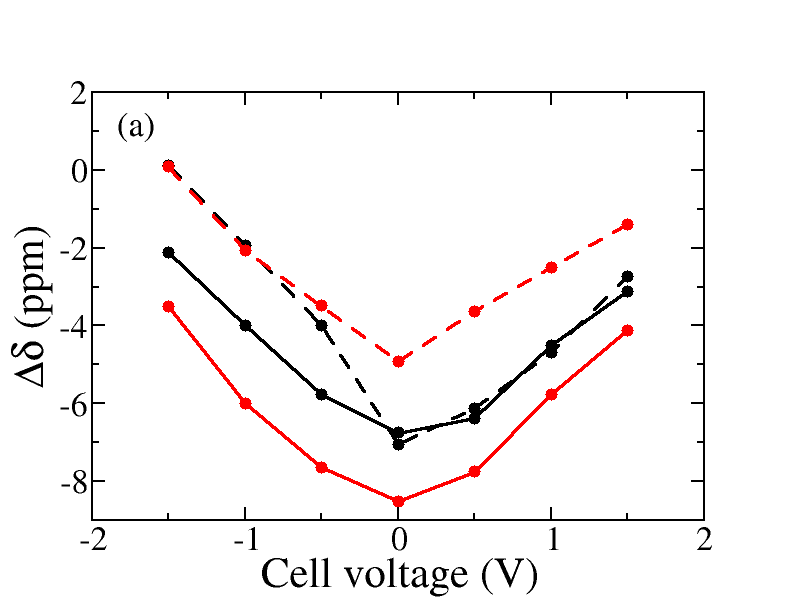}
\includegraphics[scale=0.29]{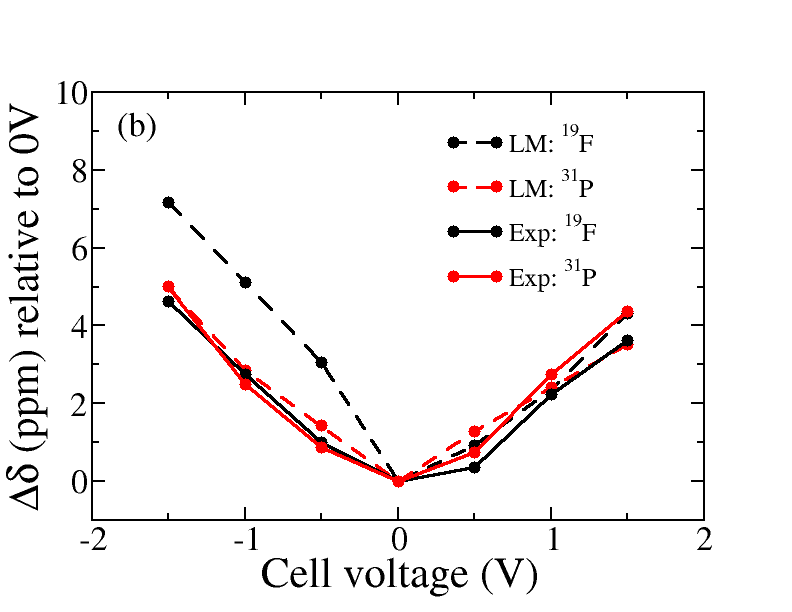}
\caption{(a) Comparison of average chemical shifts for anions and cations adsorbed inside YP-50F carbon particles relative to the bulk electrolyte ([PEt$_{4}$][BF$_{4}$] in ACN, 1.5~M) obtained through experiments~\cite{Griffin15} and lattice simulations. (b) Same as (a) but given with respect to the values at 0V.}
\label{shift_P_F}
\end{figure}

To understand the discrepancy between experiments and simulations for negative potentials, a more detailed study of the [PEt$_{4}$][BF$_{4}$]-ACN electrolyte was carried out. Figure~\ref{shift_P_F}a shows the average chemical shifts for anions and cations adsorbed inside YP-50F carbon particles relative to the bulk electrolyte ([PEt$_{4}$][BF$_{4}$] in ACN, 1.5~M) obtained through experiments~\cite{Griffin15} and lattice simulations.
For the PEt$_4^+$ cations, while the chemical shifts are underestimated by the lattice model, the variation with the potential follows closely that seen experimentally. For the BF$_4^-$ anions, while the agreement is very good on the positive side, the trend observed on the negative side diverges from the experimental one and becomes very similar to the one of PEt$_4^+$ cations. Interestingly, below -0.5~V, the lattice model gives almost identical results for BF$_4^-$ and PEt$_4^+$ indicating that the environments visited by these two ions were very similar. The excellent agreement between the trends with the applied potential difference observed experimentally and with the model is even clearer on Figure~\ref{shift_P_F}b, where chemical shifts are plotted with respect to the values at 0~V. This suggests that discrepancies between the simulations and experiments are due to inaccuracies in the representation of ion adsorption rather than in the DFT NICS calculations.  

\begin{figure}[ht!]
\centering
\includegraphics[scale=0.3]{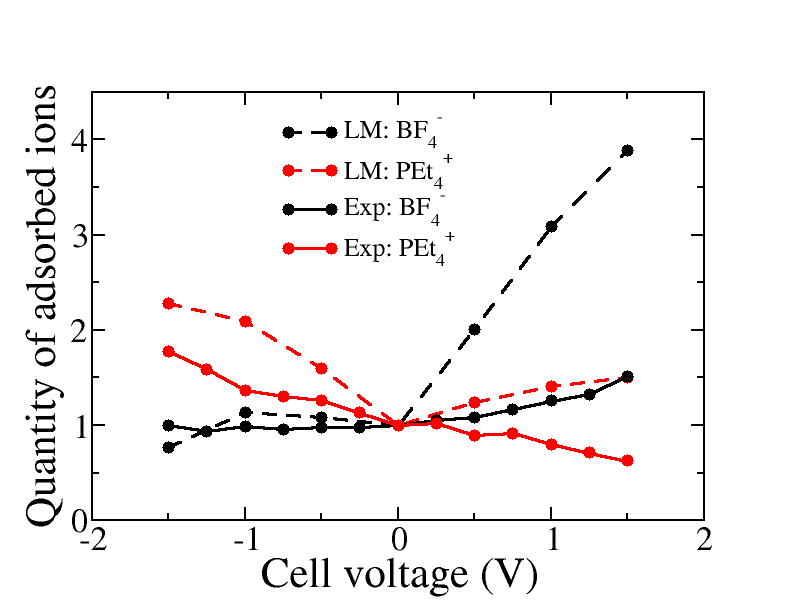}
\caption{Comparison of relative quantities of adsorbed ions determined from the lattice model and from experiments~\cite{Griffin15} for the same electrolyte at various applied potential differences with respect to the values at 0V.}
\label{shift_populations}
\end{figure}

It is important to note that while the lattice model also includes a determination of total quantities of adsorbed ions, this is not the appropriate quantity to understand the average chemical shift observed as the chemical shift depends on the distribution of ions in different environments, here different pore sizes. The quantities of adsorbed ions relative to the ones at 0~V are shown in Figure~\ref{shift_populations}.  
Considering the crudeness of the model, where the carbon particle is represented as a collection of slit pores, the overall agreement with experiment is very good. Indeed, the order of magnitude of the relative changes is well captured by the model. This plot shows, however, that the overall chemical shift does not follow the same trend as the quantity of in-pore ions. In particular, while the agreement between experiments~\cite{Griffin15} and simulations is better on the negative side, the agreement for the chemical shifts is better on the positive side. One striking feature is that the lattice model seems to overestimate the variations in quantities of adsorbed ions, especially for anions at the positive electrode. In this work, anions and cations adsorption have been considered in separate lattice simulations which can lead to an overestimation of the ion adsorption as the total ionic volume in the pores is underestimated.

\begin{figure}[ht!]
\centering
\includegraphics[scale=0.29]{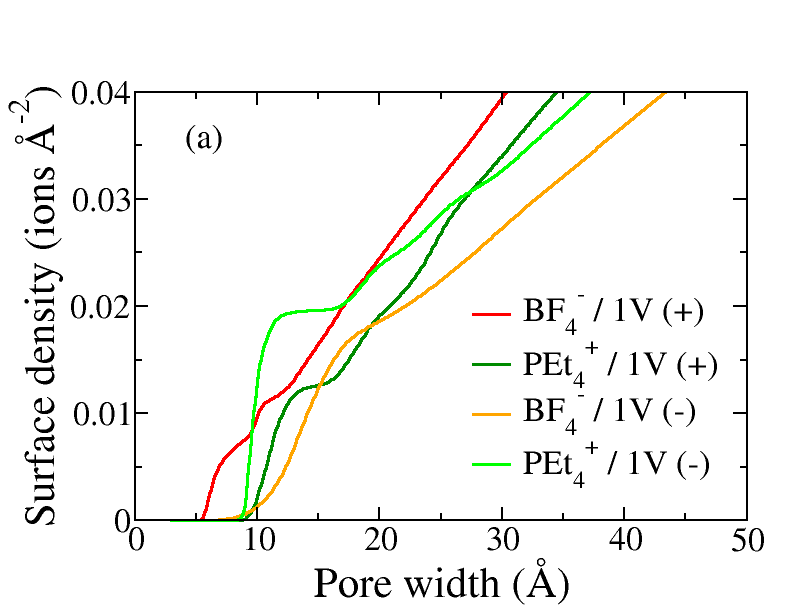}
\includegraphics[scale=0.29]{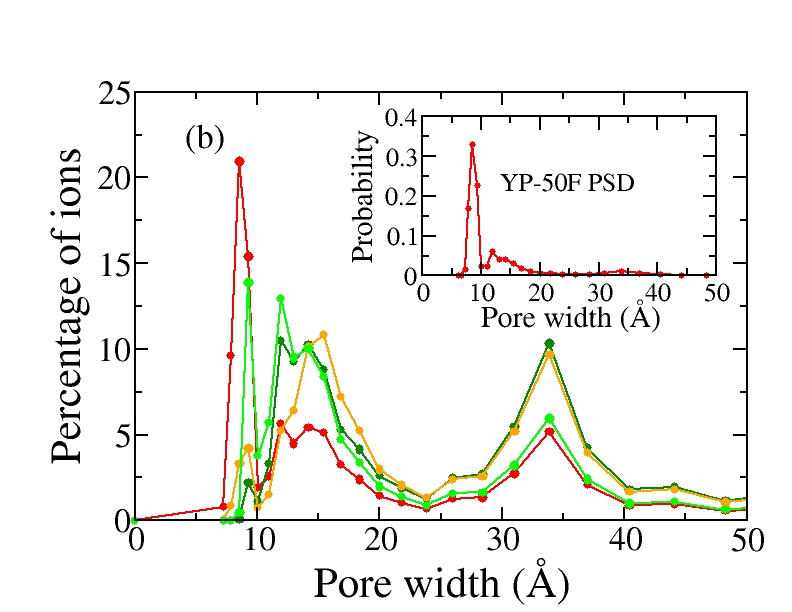}
\caption{Integrated ionic density profiles for [PEt$_{4}$][BF$_{4}$]-ACN at 1V (a) and corresponding distributions of the ions in the pores obtained with the carbon particle model (b).}
\label{Pore-pop}
\end{figure}

The environments explored by the ions are highly dependent on the surface densities of the ionic species obtained from molecular simulations results. Figure~\ref{Pore-pop}a shows the surface densities for BF$_4^-$ and PEt$_4^+$ ions at positive and negative electrodes for a potential difference of 1V. 
Apart from BF$_4^-$ at the positive electrode, the approach adopted here predicts that the surface density is zero for pores below 10~\r{A}. As the pore size distribution of YP-50F (see inset of Figure~\ref{Pore-pop}b) shows that this carbon contains a majority of pores of around 8.5~\r{A}, these low surface densities are expected to impact the actual shifts, especially as smaller pores lead to higher chemical shifts. From the lattice simulations results, it is possible to extract the distributions of ions in pores of different sizes. The distributions corresponding to a potential difference of 1~V are shown in Figure~\ref{Pore-pop}b. The pore distributions for BF$_4^-$ at the positive electrode follow quite closely the pore size distribution. This situation corresponds to a case where the agreement between simulations and experiments is excellent. It is interesting to note that for pores larger than 20~\r{A}, the relative quantities of ions are larger than the probabilities from the pore size distribution. This is expected as larger pores contain more ions. The distributions for BF$_4^-$ at the negative electrode and for PEt$_4^+$ at both electrodes indicate lower number of ions in pores below 10~\r{A}. The chemical shifts for all these situations will therefore be underestimated compared to experiments. Overall, the discrepancy between experiments and simulations in some cases, and the distributions of ions in pores of different sizes suggest that: i) the reorganisation of ions in a pore of a given size has a limited effect on the chemical shift, ii) the presence or not of ions in small pores has a large effect. Indeed, ions in small pores experience large chemical shifts in absolute values which lead to noticeable features in the experiments. In addition, the fact that the best agreement is obtained for cases where the ions are present in the pores below 10~\r{A} confirm that experimentally these pores are accessible to the ions, as expected from the large capacitances measured for this activated carbon. 

To confirm that the lattice model underestimates the quantity of adsorbed ions in small pores, additional molecular dynamics simulations were performed for [PEt$_{4}$][BF$_{4}$]-ACN (1.5~M) in contact with a porous carbon having an average pore size of 8.6~\r{A}~\cite{Deringer18}, i.e.~close to the most common pore size observed in the PSD of YP-50F (8.5~\r{A}, see Figure~\ref{Pore-pop}b). The distributions of ions in different pore sizes and the details of the molecular simulations are given in Supplementary Information. In contrast to the lattice model predictions, cations and anions both enter pores with a size below 10~\r{A} spontaneously in the molecular simulations. The actual density calculated with the lattice model for pores of 8.5~\r{A} is 1.24~10$^{-4}$~ions~\r{A}$^{-3}$ for BF$_4^-$ and 1.21~10$^{-6}$~ions~\r{A}$^{-3}$ for PEt$_4^+$. From the molecular simulations, the density is much larger, equal to 4.97~10$^{-4}$~ions~\r{A}$^{-3}$ for both anions and cations. This comparison can help explain the discrepancy between lattice simulations and experiments and shows the limit of the lattice model to predict quantities of adsorbed ions in small pores for which the use of free energy profiles extracted from simulations of an electrolyte at a planar surface is not adequate. Future works will explore the possibility of extracting three dimensional ionic densities from molecular simulations or using other theoretical methods such as molecular density functional theory.~\cite{Jeanmairet19} The presence of functional groups on the carbon surface, not taken into account in this work, and potentially leading to significant effects,~\citep{Moussa16,Dyatkin14,Dyatkin15,Dyatkin16,Dyatkin18,Lahrar21} could also be included in the future. An alternative way to exploit the lattice model would be to extract the distribution of ions in pores of different size by identifying which distribution leads to the best fit with the experimental results.

\section{Conclusion}

In this work, we have used a lattice simulation method to generate \emph{in situ} NMR spectra for various electrolyte species under a range of applied potentials. With this theoretical approach it was possible to quantify the relative importance of free energy, i.e. the ion adsorption and reorganisation, and the NICS, i.e. carbon ring currents, to the overall chemical shift. We have established that changes to the ring currents occurring in the carbon material during charging have a larger influence on the chemical shift changes. Comparisons of our model results with experimental data show that the lattice model is able to predict the variation of the chemical shifts with applied potential very well. Nevertheless, the model fails to predict the actual shift in some cases, probably as a consequence of using molecular dynamics results corresponding to electrolytes adsorbed at planar surfaces separated by distances considerably larger than the solvated ions to extract free energy profiles. The use of a realistic porous carbon electrode in molecular dynamics simulations and DFT calculations could lead to a better agreement and will be investigated in future works. Lastly, our finding that changes in ring currents dominate the observed chemical shift changes confirms that \emph{in situ} NMR spectra provide information on the electronic structure of carbon materials, providing motivation to explore this further in future studies of different carbons.

\section*{Supplementary Information}

Details of the molecular dynamics simulations, free energy profiles, details of the DFT calculations of NICS and surface charges for intrapolations, pore size and pore surface distribution along with a description of the effect of pore surface on the simulated chemical shifts.

\section*{Data availability}

The program used to do the lattice simulations is available, along with a manual, on github (https://github.com/cmerlet/LPC3D). The data corresponding to the plots reported in this paper, as well as example input files for the mesocopic model, are available in the Zenodo repository with identifier 00.0000/zenodo.0000000.

\section*{Acknowledgements}

This project has received funding from the European Research Council (ERC) under the European Union's Horizon 2020 research and innovation program (grant agreements no. 714581 and no. 835073). This work was granted access to the HPC resources of CALMIP supercomputing center under the allocations P17037 and P19003, and of TGCC under the allocations A0070911061 and A0080910463 made by GENCI. This work was also supported by a UKRI Future Leaders Fellowship to Alexander C. Forse (MR/T043024/1). The authors acknowledge Dimitrios Kilymis, El Hassane Lahrar and Dongxun Lyu for useful discussions.

\section*{Conflicts of interest}

There are no conflicts to declare.


\providecommand{\latin}[1]{#1}
\makeatletter
\providecommand{\doi}
  {\begingroup\let\do\@makeother\dospecials
  \catcode`\{=1 \catcode`\}=2 \doi@aux}
\providecommand{\doi@aux}[1]{\endgroup\texttt{#1}}
\makeatother
\providecommand*\mcitethebibliography{\thebibliography}
\csname @ifundefined\endcsname{endmcitethebibliography}
  {\let\endmcitethebibliography\endthebibliography}{}

\end{document}